# Deep learning-based reconstruction of interventional tools and devices from four X-ray projections for tomographic interventional guidance

Elias Eulig[1]  |  Joscha Maier[1]  |  Michael Knaup[1]  |  N. Robert Bennett[2]  |  Klaus Hörndler[3]  |  Adam S. Wang[2]  |  Marc Kachelrieß[1]

[1] Division of X-Ray Imaging and Computed Tomography, German Cancer Research Center (DKFZ), Heidelberg, Germany

[2] Department of Radiology, Stanford University, Stanford, California, USA

[3] Ziehm Imaging GmbH, Nürnberg, Germany

**Correspondence**
Elias Eulig, Division of X-Ray Imaging and Computed Tomography, German Cancer Research Center (DKFZ), Heidelberg, Germany.
Email: elias.eulig@dkfz.de

**Funding information**
National Center for Research Resources, Grant/Award Number: S10RR026714

**Abstract**
**Purpose:** Image guidance for minimally invasive interventions is usually performed by acquiring fluoroscopic images using a monoplanar or a biplanar C-arm system. However, the projective data provide only limited information about the spatial structure and position of interventional tools and devices such as stents, guide wires, or coils. In this work, we propose a deep learning-based pipeline for real-time tomographic (four-dimensional [4D]) interventional guidance at conventional dose levels.
**Methods:** Our pipeline is comprised of two steps. In the first one, interventional tools are extracted from four cone-beam CT projections using a deep convolutional neural network. These projections are then Feldkamp reconstructed and fed into a second network, which is trained to segment the interventional tools and devices in this highly undersampled reconstruction. Both networks are trained using simulated CT data and evaluated on both simulated data and C-arm cone-beam CT measurements of stents, coils, and guide wires.
**Results:** The pipeline is capable of reconstructing interventional tools from only four X-ray projections without the need for a patient prior. At an isotropic voxel size of 100 $\mu$m, our methods achieve a precision/recall within a 100 $\mu$m environment of the ground truth of 93%/98%, 90%/71%, and 93%/76% for guide wires, stents, and coils, respectively.
**Conclusions:** A deep learning-based approach for 4D interventional guidance is able to overcome the drawbacks of today's interventional guidance by providing full spatiotemporal (4D) information about the interventional tools at dose levels comparable to conventional fluoroscopy.

## 1 | INTRODUCTION

In the past decades, minimally invasive interventions replaced conventional surgery in many areas and enabled the development of new diagnostic and therapeutic procedures including angiography,[1–4] biopsy,[5] angioplasty using stents,[6] and embolization using coils or other emboli.[7–11] These interventions demand precise image guidance, commonly accomplished by single- or biplane X-ray fluoroscopy with typical frame rates ranging between 7.5 and 15 frames per second.[12] Such (projective) acquisitions may be supported with 3D information by occasionally acquiring cone-beam CT (CBCT) scans during the intervention.[13] However, fluoroscopy as the main source of guidance is drastically limited in its ability to resolve the three-dimensional (3D) structures and locations of the interventional material, potentially hindering the development of new procedures.[14]

Tomographic (four-dimensional) image guidance would be capable of overcoming this drawback by providing full spatiotemporal information about the interventional tools. However, to be conducted at comparable update rates as fluoroscopic image guidance and using fully sampled datasets with conventional reconstruction algorithms would result in excessively high radiation dose to both the patient and the surgeon.[15] Prior work has shown that it is possible to reconstruct interventional tools from 16 projections (without allowing for temporal overlap) using the principles of compressed sensing theory.[16,17] The PrIDICT algorithm[18,19] assumes that the rawdata difference between a forward-projected patient prior (a fully sampled patient scan acquired prior to the intervention) and the interventional data acquired at time step $t$ is zero everywhere (up to noise), except for those detector pixels, where interventional tools are present. The reconstructed volume is deteriorated in image quality due to streak artifacts coming from the high angular undersampling. To solve this problem, the authors further assume that the interventional tools are of high contrast; therefore, the insignificant voxels can be set to zero through a thresholding operation and the resulting volume is sparse, containing only the interventional tools. Adding this volume to the patient prior yields the result of the PrIDICT algorithm

$$f_t^{\text{PrIDICT}} = f_p + \theta(X^{-1}(p_t - Xf_p)), \quad (1)$$

where $f_p$ denotes the patient prior, $p_t$ the rawdata acquired during the intervention, $X$ the forward projection operator, $X^{-1}$ the Feldkamp–David–Kress (FDK)[20] reconstruction, and $\theta$ the thresholding operator. As the first assumption holds only in the absence of patient motion, it is necessary to register the patient prior before subtracting it from the interventional data. Here, a deformable volume-to-rawdata (3D-2D) registration method is preferred over a volume-to-volume (3D-3D) one in order to exclude the influence of undersampling artifacts in the image domain from the registration process.[21] However, the resulting pipeline of deformable volume-to-rawdata (3D-2D) registration method and PrIDICT algorithm[21] has two main disadvantages. First, the radiation dose levels are still approximately a factor of 16 higher than those present in today's single- or biplane fluoroscopy. Second, the registration method is too computationally intensive to realize the pipeline in real time, which would be a requirement for clinical practice.

Numerous other works proposed methods to detect and reconstruct interventional tools and devices from single- or biplane fluoroscopy data[22–27] but are limited to guide wires and catheters. A method to automatically match a 3D model of an aortic stent graft to an intraoperative 2D image showing the device has been presented for endovascular abdominal aortic repairs.[28] However, this method relies on an existing 3D model of the stent used and a series of handcrafted filters to extract the stent in the projection domain. Additional methods[29–32] have been proposed to reconstruct CT images from few projections, which are not considering interventional tools and devices but instead reconstruct patient CT images directly. Furthermore, unlike our method, some of these works[29,31] rely on prior CT scans of the same patient.

Recently, we proposed the deep tool reconstruction (DTR),[33,34] which is capable of reconstructing stents and guide wires from only four X-ray projections. Given that a patient prior is perfectly registered, this is achieved by training a Convolutional Neural Network (CNN) to learn a mapping from the corresponding sparse view CT reconstruction to a segmentation of the interventional tools. In this work, we improve this method further by first letting the network consider the 3D context rather than just two-dimensional slices and thereby improving its accuracy significantly. Second, we train the network jointly on guide wires, stents, and coils; and third, test it on scans of commercially available stents, guide wires, and coils. Furthermore, to eliminate the need for a prior volume and a computationally intensive registration step, we introduce another CNN (referred to as deep tool extraction [DTE]) in the projection domain, where interventional tools can be easily located and extracted from the background. Additionally, this eliminates the need for a patient prior, thus easing the clinical workflow and reducing the patient dose further. A similar approach, where CNNs are used to segment pacemakers in the rawdata, has recently been proposed for metal artifact reduction (MAR).[35,36] Contrary to their method, our network is trained jointly on a variety of interventional tools and devices and predicts their projection values, rather than their segmentation masks. This is necessary to both, be able to reconstruct the patient without an additional inpainting step and to reconstruct more complex objects (such as stents or coils) where parts of the tools may overlap each other in several projections.

Compared to single- or biplane X-ray fluoroscopy, our method would increase the dose by a factor four and two, respectively, if to be conducted at similar update rates. When update rates are reduced by a factor two (e.g., from 10 frames per second to 5 frames per second), this would yield twice the dose compared to single plane X-ray fluoroscopy and similar dose compared to biplane X-ray fluoroscopy. Further dose reduction possibilities, that are not focus of this paper, include the optimal choice of the tube voltage or a reduction of the tube current, for example. Moreover, there may be additional positive effects, such as not needing a prior scan or reduced beam time due to faster interventions, which may reduce the X-ray dose even further.

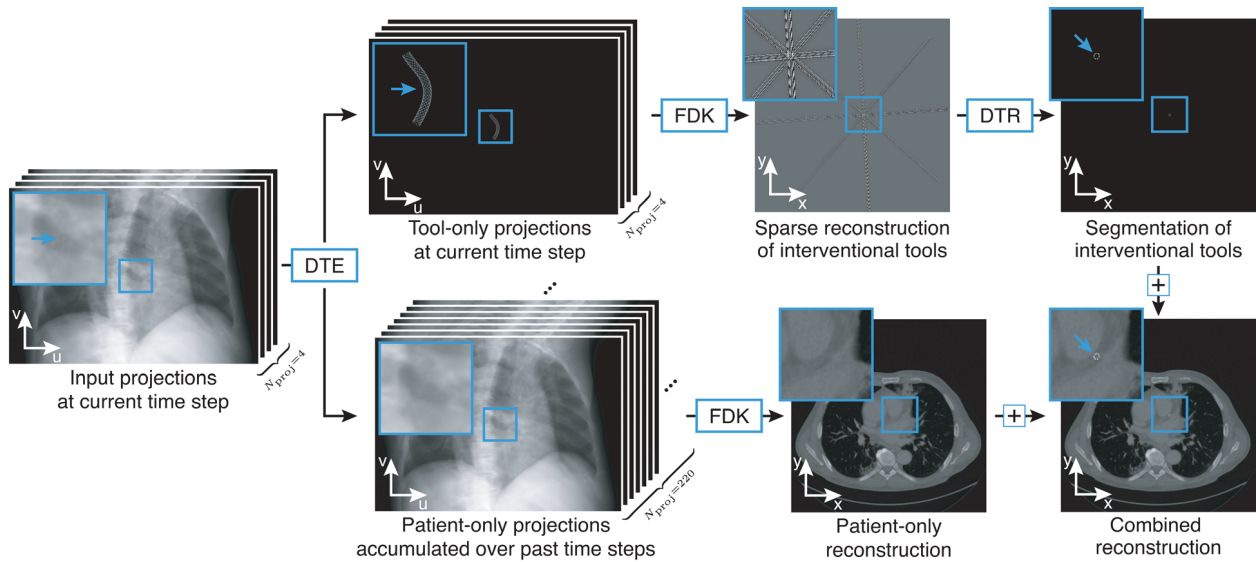

**FIGURE 1** Illustration of a deep learning-based tomographic interventional guidance. Projections are acquired continuously in which the DTE separates interventional tools from the patient background. While the reconstruction of interventional tools (upper path) is fully updated with every new set of four projections, the patient volume is reconstructed with temporal overlap using a sufficient amount of projections from distinct angles (lower path). The two volumes can be added to yield the final reconstruction

## 2 | MATERIALS AND METHODS

A novel, deep learning-based pipeline is developed, which can reconstruct the interventional tools during a procedure from only four X-ray projections and without the need for a prior scan. The method is composed of two networks, $\text{DTE}(\,\cdot\,;\epsilon)$ and $\text{DTR}(\,\cdot\,;\rho)$, where $\epsilon$, $\rho$ denote their respective set of parameters. First, the DTE network extracts the interventional tools in projections $p_t$. These projections are then FDK-reconstructed, yielding a volume with severe streak artifacts in the reconstruction that arise due to the heavy undersampling. We leverage FDK as a known operator to reconstruct using the known system geometry, as opposed to learning this transform. Then, the DTR network is applied in image domain to segment the interventional tools present in the reconstruction. During an intervention the segmentation would be added onto a continuously updated patient volume $f_p$, yielding the final volume

$$f_t = f_p + \text{DTR}\big(X^{-1}\text{DTE}(p_t;\epsilon);\rho\big). \quad (2)$$

### 2.1 | System specification

We assume a gantry comprised of four X-ray tubes and flat detectors, respectively, arranged with an angular spacing of $45°$. While the gantry rotates, new sets of projections are acquired continuously, s.t. the reconstruction of interventional tools (upper path in Figure 1) can be updated with every new set of four projections. The patient volume (lower path in Figure 1) can be FDK-reconstructed with a temporal overlap using the patient-only projections accumulated over several past time steps. Although such a system is yet to be developed, the methodology presented here is not restricted to this particular design and alterations involving fewer X-ray tubes and flat detectors potentially combined with temporally overlapping reconstructions may be made.

### 2.2 | Simulations

#### 2.2.1 | 3D models

3D models of guide wires, coils, and stents are generated through the Python API of Blender[37] with randomly varying parameters within a range such that the models reflect a wide range of commercially available tools. The resulting models are then saved as triangular meshes (STL file format) for the CT data simulation. In total, the dataset comprises 600 three-dimensional models of guide wires, stents, and coils (200 each), some of which are depicted in Figure 2.

The guide wires are simulated as cylinders with varying diameter (randomly chosen in the range of 0.4–0.8 mm based on specifications of commercial guide wires) along a nonuniform rational basis-spline (NURBS) curve with five control points, evenly spaced along the $z$-axis. To deform the curve, between two and four points are randomly shifted in the $x$–$y$ plane based on a multivariate normal distribution with $\mu = 0$ and $\sigma = 2.5$ cm. The resulting models are then randomly rotated along all three axes, to ensure that the training data are not biased toward a specific orientation of the guide wires.

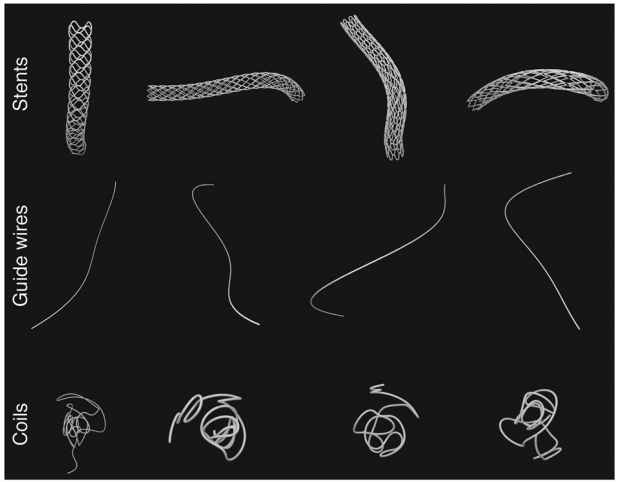

**FIGURE 2**  Some examples for 3D models of stents, coils, and guide wires used to train and validate both DTE and DTR

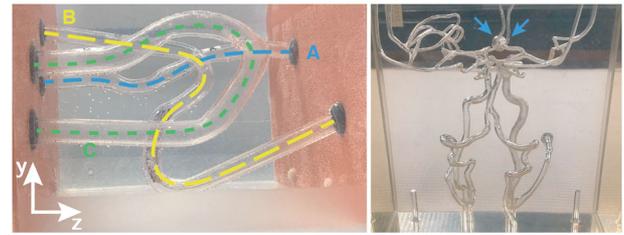

**FIGURE 3**  Custom vessel phantom (left) and commercial brain-vessel phantom (right) with a saccular aneurysm (blue arrows). The scans acquired using these phantoms were used to test the proposed pipeline

Each stent model is based on one out of eight base models, with varying stent diameters and strut thicknesses. The diameters are randomly chosen in the range from 4.8 to 6.3 mm and the strut thicknesses are randomly chosen in the range from 0.4 to 0.7 mm based on specifications of commercial stents. The resulting models are placed along a similar NURBS curve as the guide wires and between one and three of the control points are randomly shifted in the x–y plane based on a multivariate normal distribution with $\mu = 0$ and $\sigma = 1$ cm. Finally, the models are randomly rotated along each of the x-, y-, and z-axes.

The coil models are generated by placing between 30 and 60 control points within a sphere of a diameter between 10 and 30 mm. The control points are placed equidistantly from each other and with a minimum opening angle of 45° to prevent unrealistically strong kinks. Then, the points are connected through a NURBS curve with each other. A cylinder with diameter between 0.4 and 0.8 mm following this curve yields the final 3D model.

### 2.2.2  |  CT data simulation

To simulate cone-beam projections, the simulated tools are randomly placed within the field of measurement (FOM) of the scanner and polychromatically forward-projected according to the Zeego geometry. The material of the tools is simulated to be iron of varying density (between 3.2 and 3.9 g/cm$^3$), such that the resulting absorption coefficients match the ones of commercially available tools. For the training of the DTE network, 17 projections per 3D model were simulated with an angular spacing of 10° and the resulting projections were cropped around the interventional tools prior to loading them into the training pipeline described in Section 2.4.

For the training of the DTR, the models are forward projected from four projections with an angular spacing of 45° and a random start angle. The volume around the tools is reconstructed using a FDK reconstruction with an isotropic voxel size of 100 $\mu$m$^3$. The respective voxel representations of the STL files serve as ground truth.

### 2.3  |  Measurements

Ideally, our method would be tested on CT scans from real interventions, together with a ground truth segmentation of the interventional tools. However, CBCTs are rarely being acquired during interventions and for these an acceptable segmentation is often not available. Hence, to obtain a good estimate of the potential performance on clinical data, the method is tested on phantom measurements instead.

To this end, scans of interventional tools, namely, guide wires, stents, and a coil are acquired using two different intervention phantoms. For the measurement of stents and guide wires, a vessel phantom (Figure 3) is designed and constructed, consisting of artificial vessels in a water container, in which the stents and guide wires can be inserted. They are then scanned in a variety of different positions, leading to different deformations of the tools. For the measurements of the coil, a commercial brain vessel phantom of the circle of Willis is used (Elastrat Sàrl, Geneva, Switzerland). The coil is inserted in a saccular aneurysm at the right A1 segment of the anterior cerebral artery (Figure 3, blue arrows) and three scans are acquired at different insertion depths.

To obtain the ground truth, we acquired prior scans of the phantom only and subtracted the projections from the interventional ones for all 496 projections. Thresholding after the FDK reconstruction yields the desired ground truth segmentation.

For all measurements of interventional tools and devices, a Zeego robot-driven C-arm system (Siemens Healthineers, Forchheim, Germany) with specifications given in Table 1 is employed.

**TABLE 1** Overview of system specifications of the Zeego robot-driven C-arm system used in this study

| Specification | Description |
|---|---|
| Focal-spot-to-detector distance | 1200 mm |
| Focal-spot-to-isocenter distance | 785 mm |
| Detector size | 40 cm × 30 cm |
| Detector material | a-Si with CsI(Ti) scintillator |
| Pixel size $d_{uv}$ | 308 µm |
| Matrix size $N_u \times N_v$ | 1240 × 960 |
| # of projections over $\pi$ + fan angle | 496 |
| Field of measurement (FOM) | 25 cm |
| Tube voltage | 90 kV |

## 2.4 | Deep tool extraction

To eliminate the need for both a patient prior and a computationally intensive registration algorithm as it was the case in prior work,[18,19,21] the interventional tools are extracted directly in the raw data domain using a CNN. From an arbitrary projection showing interventional tools in a patient, this network is trained to predict the projection values of the interventional tools.

### 2.4.1 | Data preprocessing

For training and validation of the network, simulated forward projections of interventional tools that are added onto projections of CBCT patient scans serve as input and the respective projections of interventional tools serve as ground truth. Projections from four CBCT patient scans (two thorax, one head, and one pelvis) are used for training and the projections from two CBCT patient scans (one thorax and one abdomen) are used for validation. Training and validation is performed patch-wise on patches of size 384×384 pixels that are randomly cropped from random projections of the patient scans. To reduce the amount of training data with interventional tools placed outside the patient, we reject those patches with no or minor attenuation. The projections of the tools are then randomly flipped in x–y direction to cover the remaining 180°, which were not simulated and are added to the patient patch in a random position.

To let the network generalize to different noise levels, we simulate quantum noise using a Poisson distribution for varying tube currents. Furthermore, to let the network generalize to different point spread functions (PSFs), angular blurring from the rotation of the scanner, and small movements of the interventional tools, the projections are blurred with a multivariate Gaussian kernel with standard deviations sampled uniformly between 0 and 1.4 pixels. Contrast media was neither present in the patient data, nor simulated by us. Nevertheless, in a prior study, we had demonstrated that contrasted vessels can very well be extracted from fluoroscopic images by similar means.[38,39]

As all data preprocessing is performed in an online manner, we estimate the mean and standard deviation of projection values using a running mean and running variance.[40,41] Each sample is then normalized with the current statistics, such that the training data are normalized to zero mean and unit variance at any time.

### 2.4.2 | Training details

As network $DTE_\epsilon$, we use a standard U-Net[42] composed of four stages in both the encoding path and the decoding path. Here, each stage is composed of two convolutional layers (kernel size = 3×3, stride = 1×1, zero padding) each followed by a batch normalization layer,[43] a spatial dropout layer[44] with dropout probability 0.2 to reduce overfitting and a rectified linear unit (ReLU).[45] The encoding path performs a progressive spatial downsampling by employing 2×2 max-pooling after each stage while increasing the feature dimension by a factor of 2 starting with 64 in the first layer. The upsampling in the decoding path is performed by a nearest-neighbor upsampling operation followed by a convolutional layer (kernel size = 3×3, stride = 1×1, zero padding). Concatenative skip connections between the encoding and the decoding path allow high-resolution information from the earlier layers to skip the bottleneck. All convolutional layers of the network are initialized using He initialization,[46] whereas the weights of the batch normalization layers are initialized with unity and their biases with zeros.

The training pipeline is implemented using PyTorch.[47] The optimal parameters $\hat{\epsilon}$ are determined by minimizing the $\mathcal{L}_1$ loss between network output $DTE(x; \epsilon)$ and ground truth $y$ on mini-batches of size $N = 6$

$$\hat{\epsilon} = \arg\min_{\epsilon} \frac{1}{N \times K^2} \sum_{n,k} \|DTE(x; \epsilon)_{n,k} - y_{n,k}\|_1, \quad (3)$$

where the sum runs over all samples $0 \leq n < N$ and pixels $0 \leq k < K^2$, respectively. To this end, the Adam optimizer[48] with a learning rate $\alpha = 1 \times 10^{-4}$ and parameters $\beta_1 = 0.9$, $\beta_2 = 0.999$ is employed.

## 2.5 | Deep tool reconstruction

The second stage of our pipeline takes as input several sparse FDK reconstructions of tool-only projections that are obtained by the DTE. By doing so, DTR is able to learn 3D features about the structure of the interventional tools. The DTR is then trained to predict the segmentation of the corresponding interventional tools.

### 2.5.1 | Data preprocessing

During training, the network takes as input chunks of $C$ patches of size 224×224 pixels, with $C$ as an uneven number of reconstructed slices. These chunks are randomly cropped from the FDK reconstruction together with the ground truth of the center slice. Prior to feeding the reconstructions to the network, they are normalized to have zero mean and unit variance.

### 2.5.2 | Training details

The DTR network is very similar to the DTE network, with two differences. First, the multichannel input to the network are the aforementioned $C$ $x$–$y$ slices of the sparse FDK reconstruction. Second, for the final layer of the DTR network, we employ a sigmoid, rather than an ReLU as nonlinearity.

The training pipeline is implemented using PyTorch.[47] Training is performed with a mini-batch size of $N = 16$ and we employ Adam[48] ($\alpha = 1 \times 10^{-4}$, $\beta_1 = 0.9$, $\beta_2 = 0.999$) to minimize the soft Dice loss with Laplace smoothing between the ground truth segmentation of the center slice $y$ and the respective network output $\text{DTR}(x; \rho)$,

$$\hat{\rho} = \arg\min_{\rho} \left[ 1 - \frac{2 \sum_{n,k} \text{DTR}(x;\rho)_{n,k} y_{n,k} + 1}{\sum_{n,k} \text{DTR}(x;\rho)_{n,k} + \sum_{n,k} y_{n,k} + 1} \right]. \quad (4)$$

To obtain a segmentation during inference, we threshold the probabilistic output of the network such that an output pixel value is rounded in order to yield the values 0 or 1, with 0 being a classification as background and 1 being a classification as an interventional tool.

## 2.6 | Combined pipeline

The combined pipeline is composed of the following steps. First, four CBCT projections with an angular spacing of 45° are fed successively to the DTE, which extracts the tools from the patient anatomy. The volume is then sparsely reconstructed using the FDK reconstruction. The DTR network is applied to the sparse reconstruction by successively feeding chunks of consecutive slices to the network, leading to the segmentation of interventional tools present within the volume. This volume can be 3D rendered[37] for visualization purposes.

## 3 | RESULTS

### 3.1 | Deep tool extraction

The DTE is trained and validated using simulated projection data as described in Section 2.4. Figure 4 provides nine exemplary samples from the validation data together with the respective DTE prediction and ground truth. For all three interventional tools, we observed little deviation from the ground truth. In some of the samples (blue arrows), three small structures can be noticed, which were detected by the network as being interventional material and do not belong to the simulated tools. A further investigation of the corresponding patient scan revealed that these structures are indeed surgical metal clips and were thus correctly detected by the network as interventional material. This is a strong indicator that the network generalized well, beyond those interventional tools seen during training.

The simulation involved in the training and validation of the DTE does not place the interventional tools in a physiological meaningful way inside the patient and furthermore neglects several physical effects such as scatter. To analyze the method's performance in a clinical setting, we therefore applied the DTE to three fluoroscopy scans that were acquired using a Ziehm Vision RFD C-arm system (Ziehm Imaging, Nuremberg, Germany) with a pixel size of 288 µm at 90 kV tube voltage during minimally invasive angioplasties with stenting (Figure 5).

We notice significant deviation of the tools present in the patients from the tools which we had simulated as well as contrast media present in all three patients (clearly perceptible in Patient 3). Nonetheless, DTE detects most of the interventional tools present in the data, with only few exceptions such as few struts of some of the stents or the catheter in the second patient scan. The method's ability to detect radiopaque markers on the stents and guide wires is particularly remarkable, considering such markers were not included in the simulated data. Due to the lack of ground truth data for these scans, the analysis remains qualitatively.

### 3.2 | Deep tool reconstruction

The DTR was trained to segment interventional tools in volumes that were sparsely reconstructed from only four X-ray projections. It is assumed that these projections contain only noise-free projection values of the interventional tools, which are perfectly extracted from the patient anatomy.

We investigated the effect of the number of input slices $C$ on the network's performance by evaluating the Sørensen–Dice coefficient (Dice score), precision, and recall on the validation dataset for different numbers of input slices (Table 2).

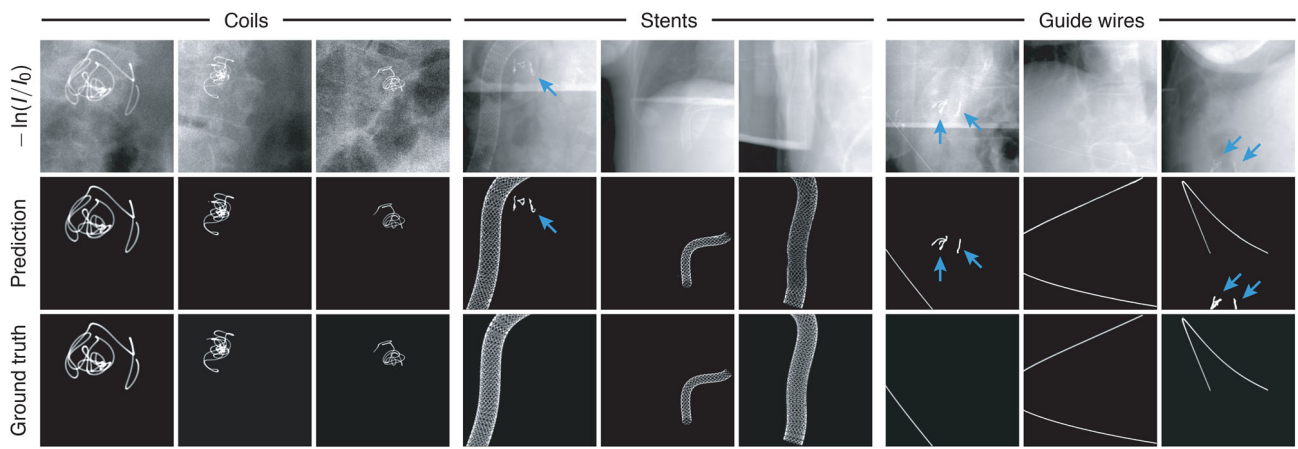

**FIGURE 4** Application of the deep tool extraction to nine exemplary samples from the validation dataset. All patches were normalized individually for visualization purposes and prediction and ground truth patches were normalized identically for all samples

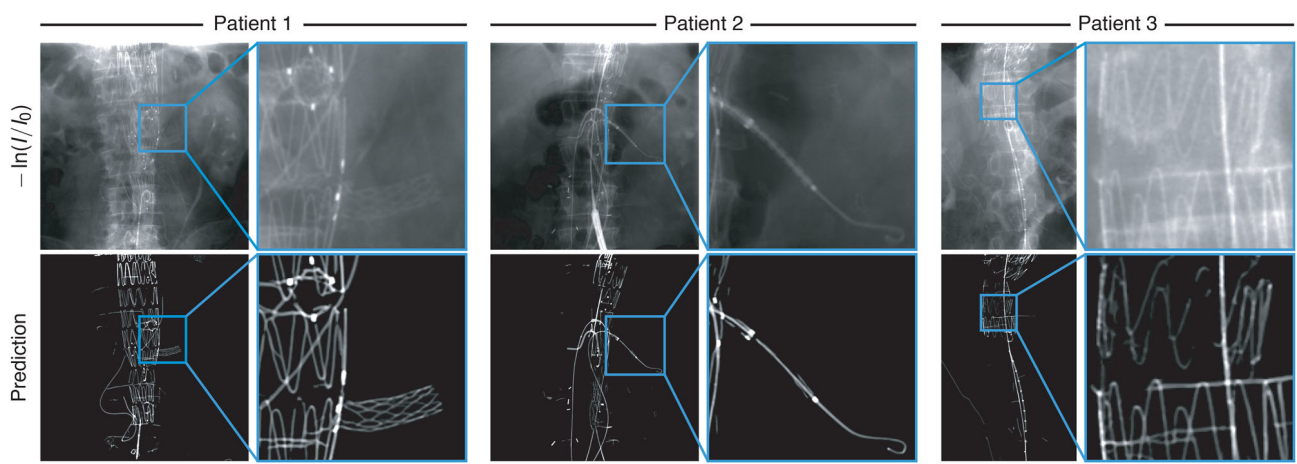

**FIGURE 5** Application of the deep tool extraction to three contrast-enhanced patient scans, which were acquired using a Ziehm Vision RFD C-arm system with a tube voltage of 80 kV during a minimally invasive angioplasty with stenting. Image courtesy of Ziehm Imaging, Nuremberg, Germany

It can be concluded that the segmentation performance increases with an increasing number of input slices. This is expected due to the larger receptive field in $z$-direction associated therewith. However, this improvement is significant only for few slices, and we cannot observe a significant improvement for $C > 5 \cong 0.5$ mm.

In Figure 6, we provide nine exemplary samples from the validation data together with the respective DTR segmentation and ground truth. For all samples shown, only little to no deviation from the ground truth was observed (as indicated by the blue arrows).

### 3.3 | Combined pipeline

Both submethods were combined and applied to measurements of commercially available interventional tools (Section 2.3). As these measurements were performed in phantoms, extracting the tools from the background is a trivial task, and consequently would lead to an overestimation of the performance of both DTE and DTR. Instead, the rawdata difference of interventional scan (phantom + tool) and prior (phantom

**TABLE 2** DTR segmentation results evaluated using Dice score, precision (positive predictive value), and recall (sensitivity) for different numbers of input slices. Highlighted are the maximum values for each metric across all configurations

| C | z-coverage | Dice score | Precision | Recall |
|---|---|---|---|---|
| 1 | 0.1 mm | 0.751 ± 0.075 | 0.761 ± 0.056 | 0.745 ± 0.101 |
| 3 | 0.3 mm | 0.781 ± 0.070 | 0.797 ± 0.047 | 0.769 ± 0.095 |
| 5 | 0.5 mm | **0.794 ± 0.065** | **0.803 ± 0.047** | **0.790 ± 0.087** |

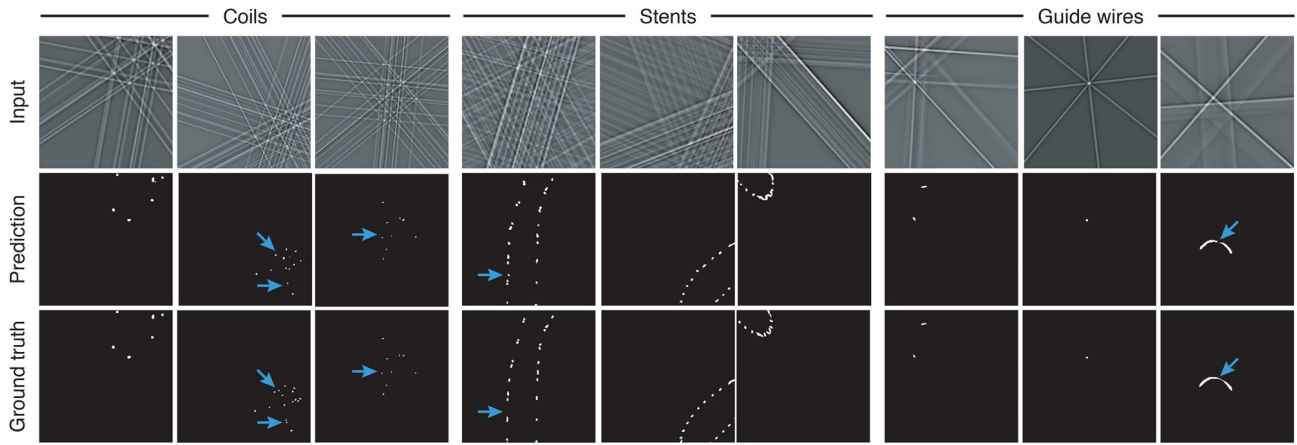

**FIGURE 6** Application of the deep tool reconstruction to nine exemplary samples from the validation dataset. All patches were normalized individually for visualization purposes and prediction and ground truth patches were normalized identically for all samples

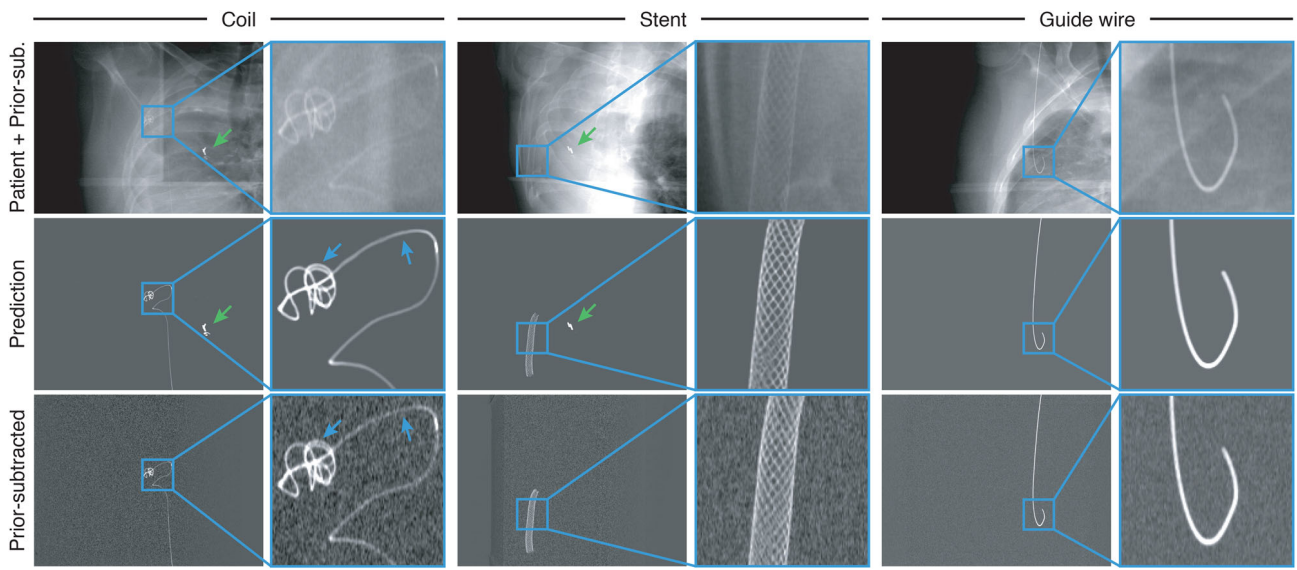

**FIGURE 7** Application of the deep tool extraction to phantom scans that were superimposed with a patient scan. Note that the noise distribution of the prior-subtracted scan is not corresponding to the projection values of the patient scan. The radiopaque markers (green arrows) are part of the patient scan and are therefore not visible in the prior-subtracted images

only) was superimposed with a thorax patient scan, which was neither part of the training nor of the validation dataset for the DTE. The projection values of these projections were then normalized with the mean and standard deviation obtained during the training of the DTE prior to feeding them into the network.

Results of applying the DTE to scans of a stent, guide wire, and coil are shown in Figure 7. It was found that not only did DTE successfully extract the tools from the background, but it even enhanced the resolution of fine details such as the individual struts of the stent or parts of the coil (blue arrows). This can be explained by the DTE having been trained on simulated data, where the ground truth is noise-free and not degraded by blurring caused by the PSF of the system or by a misalignment between the interventional scan and the prior scan. Furthermore, DTE detected two radiopaque markers (green arrows) as interventional tools, even though radiopaque markers were never included in the training data. In Table 3, we report the mean absolute percentage error (MAPE) for the detector pixels, where a tool is present in the prior-subtracted scan. Note that, to reduce the influence of noise present in the prior-subtracted scan, we filtered both the DTE output and the prior-subtracted scan with a 3×3 pixels median filter.

**TABLE 3** MAPE of DTE, evaluated on detector pixels where tools are present in the prior-subtracted scan

| Tool | MAPE [%] |
| --- | --- |
| Guide wires | 6.0 ± 0.1 |
| Stents | 13.4 ± 2.1 |
| Coils | 13.2 ± 1.6 |

Overall, little deviations from the prior-subtracted scan were observed, and some of these deviations may be explained by the prior-subtracted scan being degraded by noise. Together with the results seen in Figure 5, this shows that the DTE generalized well beyond simulated data and can successfully extract interventional tools from background anatomy.

The combined pipeline was applied to scans of two stents (Figure 8), the coil (Figure 9), and two guide wires (Figure 10). For the stents, scans of different positions inside the phantom, leading to various deformations of the stents, are provided. For the coil, three scans at different insertion depths into the aneurysm are shown.

A very good accordance with the ground truth was found for the stent reconstructions. This can be observed particularly well, when comparing the 3D renderings with each other. Most deviations were caused by split errors, where the DTR segmentation split one single intersection in the ground truth into two separate ones (see blue arrows). Furthermore, in the reconstruction of the second position of the second stent, a slightly worse performance can be observed, most likely resulting from the stent being partially oriented in the $x$–$y$ plane. In such cases, a higher number of intersection points per plane occurs, making the reconstruction more ambiguous. The accuracy for such cases could be improved by increasing the number of projections, thereby reducing the number of possible solutions.

A similarly good performance as for the stent reconstructions can be observed for the coil reconstructions. For the scan where the coil is fully inserted into the aneurysm, however, we observe some deviations from the ground truth, specifically right after the coil leaves the micro catheter (blue arrows). A closer inspection of the DTE output revealed that some of these parts were not detected in one of the projections by the DTE and therefore could not be reconstructed by the DTR that followed.

For the guide wires, almost no deviations from the ground truth were found. This is expected due to their low number (usually less than four) of intersections with a given reconstructed slice. From discrete tomography, it is well known that any slice with $n$ intersections is uniquely determined by $m > n$ projections from mutually nonparallel directions,[49–51] which is often the case for guide wires but usually not for coils or stents. We report the Dice score, precision and recall on the stents, guide wires, and the coil in Table 4. To accommodate for the fact that a pixel-perfect segmentation with 100 μm resolution would not be of practical interest, we furthermore consider the precision $P_{o,100\,\mu m}$ and recall $R_{o,100\,\mu m}$ within a 100 μm ($\hat{=}$ 1px) environment of the ground truth segmentation, via dilation/erosion of said ground truth.

### 3.4 | Reducing the number of projections

To investigate the effect of fewer X-ray tubes (and detectors) on the performance of the DTR, we conducted additional studies with two and three projections, respectively. For this, we kept the total angular range constant (at 180° as for the previous experiments) in order to not increase the amount of limited angle artifacts present in the scans. The DTR networks trained on simulated data with two/three projections were then applied to sparse reconstructions from two/three projections, which were preprocessed by the DTE network. A quantitative comparison of reconstruction quality using different numbers of projections is given in Figure 11 and exemplar reconstructions of one guide wire, one stent, and one coil using different numbers of projections are shown in Figure 12.

Although the reconstruction quality is still consistently high across all tools and devices (see Figure 11) when reconstructing using three projections, we find a significant degradation in reconstruction quality when reconstructing from two projections. This is particularly the case for coils, where not only the number of intersections per plane is usually high (compared to guide wires), but also constraints on those intersections are weak and thus the amount of prior knowledge that may be learned by the network is low (compared to stents).

### 3.5 | Reducing the angular distance between adjacent projections

Furthermore, to investigate the effect of smaller angular distances between adjacent projections on the performance of the DTR network, we conducted additional studies with angular distances of 30° and 22.5°, respectively. The number of projections was kept constant at four.

Here, we find that while a degradation of reconstruction quality can be observed for smaller angular distances between adjacent projections (Figure 13), the overall reconstruction quality remains high even with 22.5° for all tools and devices (Figure 12).

## 4 | DISCUSSION

The complete pipeline has few limitations in its current state that we intend to investigate in future studies. A gantry with four rotating flat detectors and X-ray tubes is yet to be developed and it is unlikely that such a system will be produced in the near future due to the high associated costs. However, our experimental results with three and two projections showed that reconstruction quality remains high for all tools and devices considered in this work when reconstructing from three projections. Furthermore, it may be possible to reconstruct the volumes with a temporal overlap, where the four projections are not acquired simultaneously and thus less than four X-ray tubes and flat detectors would be required. However, this would be possible only if the motion of the tools was either negligible or accounted for by the pipeline by incorporating motion correction. Moreover, we focused on a standard 2D U-net architecture and simple learning schemes for both DTE and DTR. Although some experiments involving attention-gated U-nets[52] were conducted, no significant advantage over a standard U-net was found. Nonetheless, more advanced network architectures and learning schemes are likely to improve the proposed method further. Particularly, leveraging the temporal information using a recurrent neural network based on long short-term memory[53–55] or gated recurrent units[56] and utilizing higher level features using generative adversarial networks[57,58] could be very promising approaches for future studies. Another possible improvement could be made by jointly training the DTE and DTR networks end-to-end, thereby allowing the DTR network to account for possible errors made by the DTE. Lastly, this work did investigate the generalization capabilities of our approach only to a limited extent. Our experiments showed that the method can generalize to patient data including a variety of interventional tools and devices that were acquired at different noise levels. Although we did not account for contrast media in the patient scans during training, we found indication that the DTE is able to extract interventional tools and devices in the presence of contrast media. However, a thorough evaluation of the generalizability of our approach regarding contrast media, more diverse interventional tools and devices, different noise levels, and different scanners remains future work.

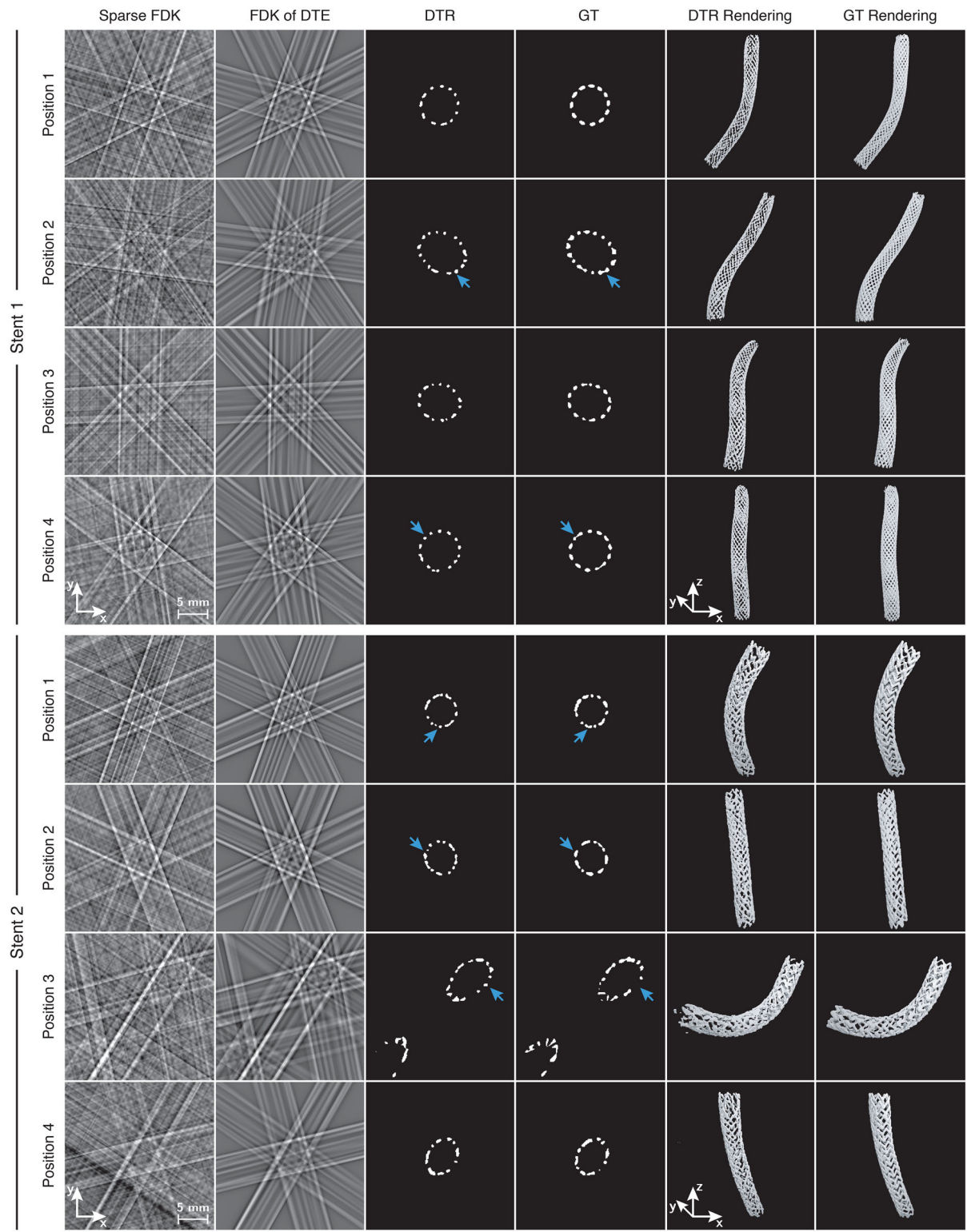

**FIGURE 8** Application of the combined pipeline to scans of the first test stent (top) and second test stent (bottom). The rows correspond to different positions of the stents inside the vessel phantom. All slices are 224×224 pixels subsets

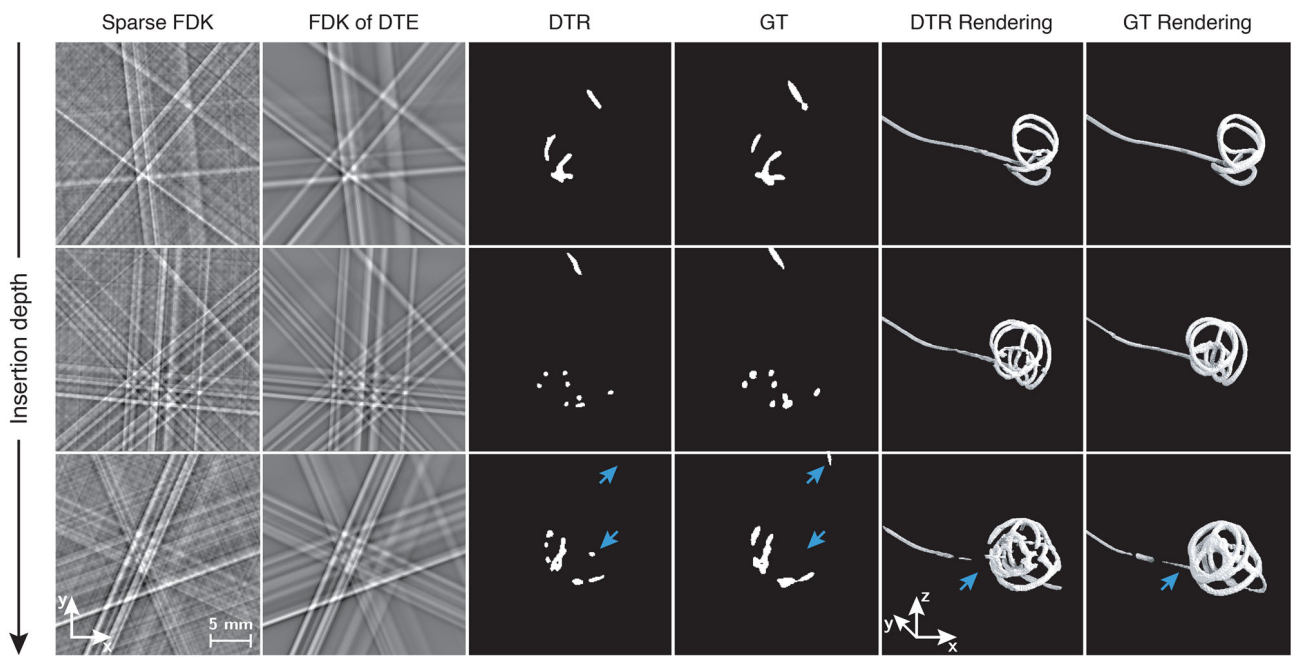

**FIGURE 9** Application of the combined pipeline to three scans of the coil at different insertion depths into the aneurysm. All slices are 224×224 pixels subsets

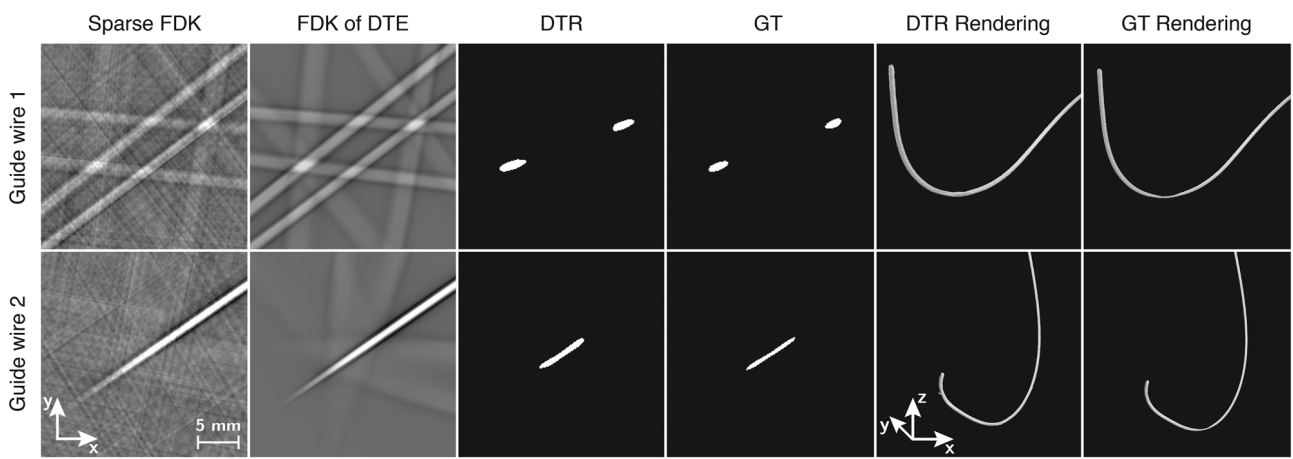

**FIGURE 10** Application of the combined pipeline to scans of two guide wires. All slices are 224×224 pixels subsets

**TABLE 4** Segmentation results of the combined pipeline applied to the test data evaluated using Dice score, precision, recall, and precision and recall ($P_{o,100\mu m}$, $R_{o,100\mu m}$) in a 100 μm environment of the ground truth

| Tool | Dice score | Precision | Recall | $P_{o,100\mu m}$ | $R_{o,100\mu m}$ |
|---|---|---|---|---|---|
| Guide wires | 0.85 ±0.03 | 0.80 ± 0.09 | 0.91 ± 0.06 | 0.93 ± 0.05 | 0.98 ± 0.02 |
| Stents | 0.63 ± 0.04 | 0.76 ± 0.06 | 0.55 ± 0.05 | 0.90 ± 0.05 | 0.71 ± 0.06 |
| Coils | 0.69 ± 0.04 | 0.85 ± 0.02 | 0.58 ± 0.05 | 0.93 ± 0.01 | 0.76 ± 0.06 |

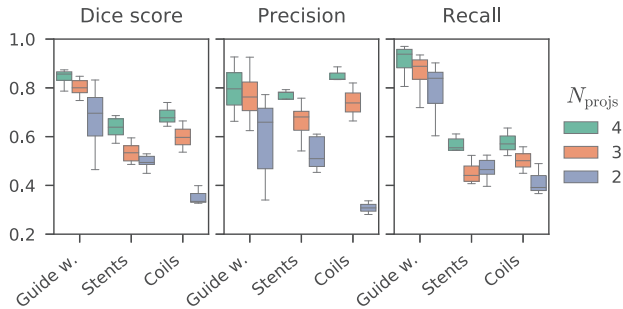

**FIGURE 11** Dice score, precision, and recall for all tools and devices when reconstructing using $N_{\text{projs}} = \{4, 3, 2\}$ projections

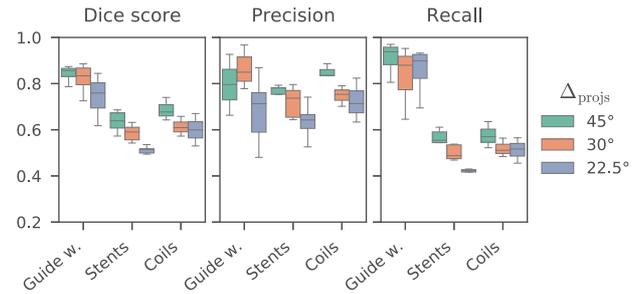

**FIGURE 13** Dice score, precision, and recall for all tools and devices when reconstructing using $\Delta_{\text{projs}} = \{45°, 30°, 22.5°\}$

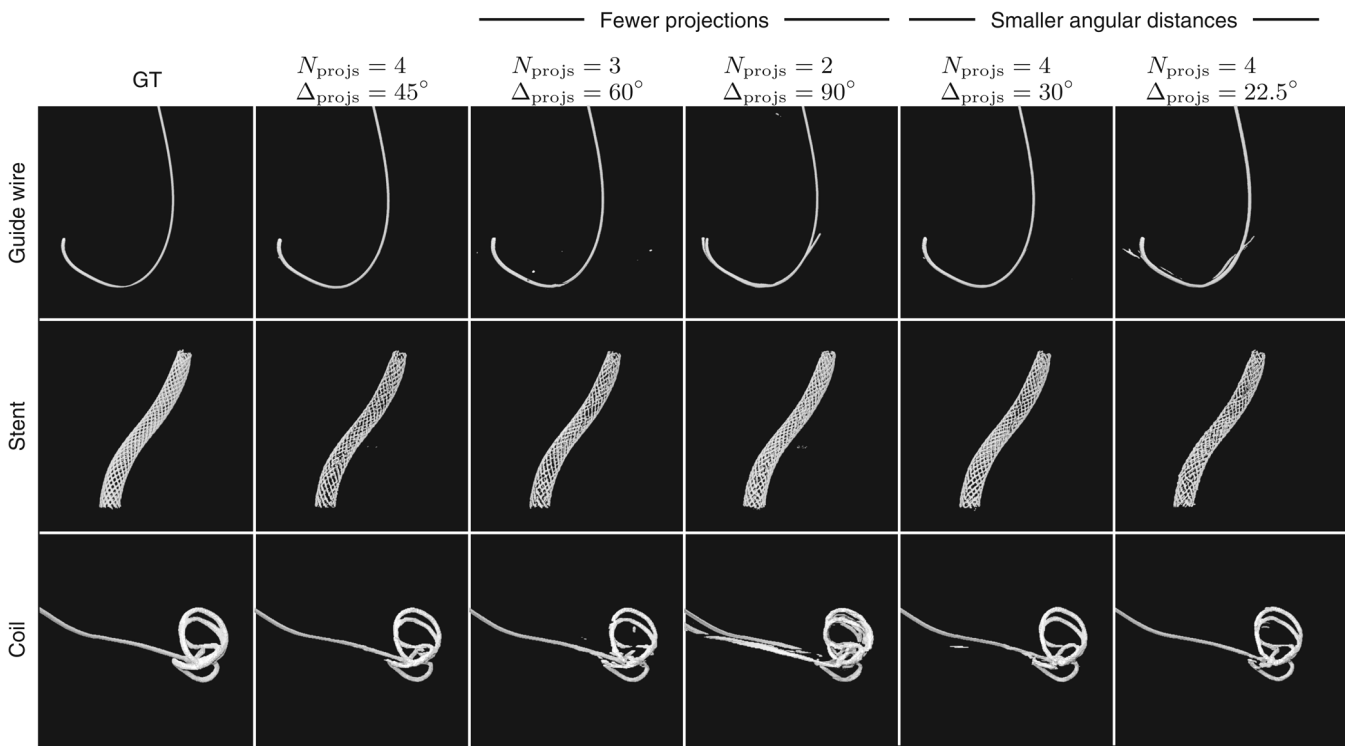

**FIGURE 12** Ground truth reconstructions of three exemplar tools and devices, and their reconstructions using the default parameters ($N_{\text{projs}} = 4$, $\Delta_{\text{projs}} = 45°$) and using fewer projections (Section 3.4) and smaller angular distances (Section 3.5)

## 5 | CONCLUSIONS

The proposed algorithm is capable of reconstructing interventional tools and devices from only four CBCT X-ray projections without the need for a patient prior. To our knowledge, this is the first time that CNNs were applied to realize tomographic interventional guidance and the results indicate that our method could overcome the current drawbacks of conventional interventional guidance, which is 2D fluoroscopy or 3D static tomography, by providing the surgeon with full real-time spatiotemporal information about the location and structure of interventional tools and devices during a minimally invasive intervention at similar X-ray dose levels. Not only could this enable the development of new procedures but it could also reduce the risk of complications for already existing procedures.

**ACKNOWLEDGMENTS**
Parts of the reconstruction and simulation software were provided by RayConStruct® GmbH, Nürnberg, Germany.

The study was supported in parts by the Society of High Performance Computational Imaging (SHPCI) e.V., Nürnberg, Germany, by the National Center for Research Resources of the National Institutes of Health under award number S10RR026714, and by Siemens Healthineers.

Some of the computing for this project was performed on the Sherlock cluster. We would like to thank Stanford University and the Stanford Research Computing Center for providing computational resources and support that contributed to these research results.

Open Access funding enabled and organized by Projekt DEAL.

**CONFLICT OF INTEREST**
Klaus Hörndler is managing director at Ziehm Imaging GmbH.

**DATA AVAILABILITY STATEMENT**
Research data are not shared.